\documentclass[preprint,floatfix] {revtex4} 
 
\usepackage{graphicx}
\usepackage{subfigure}
\begin{document}

\title{Bound states of the generalized spiked harmonic oscillator}
\author{Amlan K. Roy}
\altaffiliation{Corresponding author. Email: akroy@chem.ucla.edu. Present address: Department of 
Chemistry, University of Kansas, Lawrence, KS, 66045, USA.}
\affiliation{Department of Chemistry and Biochemistry, University of California, 
Los Angeles, CA, 90095-1569, USA}

\author{Abraham F. Jalbout}
\affiliation{Institute of Chemistry, National Autonomous University of Mexico,
Mexico City, Mexico}

\begin{abstract}
Bound states of the generalized spiked harmonic oscillator potential are calculated accurately by 
using the generalized pseudospectral method. Energy eigenvalues, various expectation values, 
radial densities are obtained through a nonuniform, optimal spatial discretization of the radial 
Schr\"odinger equation efficiently. Ground and excited states corresponding to arbitrary values of 
$n$ and $\ell$ are reported for potential parameters covering a wide range of interaction. Both weak and 
strong coupling is considered. The effect of potential parameters on eigenvalues and densities 
are discussed. Almost all of the results are reported here for the first time, which could be useful 
for future studies.
\end{abstract}
\maketitle

\section{Introduction}
The spiked harmonic oscillator (SHO) defined by the Hamiltonian $H=-d^2/dr^2 +r^2+\lambda 
r^{-\alpha}$, where $r \in [0,\infty]$ and $\lambda$, $\alpha$ are positive definite parameters 
signifying the strength of perturbation and type of singularity, has been of considerable 
interest for over three decades. This has a number of interesting physical as well as 
mathematical properties and found applications in chemical, nuclear and particle physics. For 
example, this represents a prototype of the so-called Klauder phenomenon \cite{detwiler75,
klauder78}, i.e., even after the perturbation is completely turned off ($\lambda \rightarrow 0$), 
some interaction still remains. The Rayleigh-Schr\"odinger perturbation series diverges following 
the relation $n\ge 1/(\alpha -2)$, with $n$ denoting the order of perturbation. Another interesting 
feature is that in the region of $\alpha \ge 5/2$, this gives rise to supersingularity, i.e., the 
potential is so singular that any nontrivial correction to the energy (matrix element) diverges 
\cite{detwiler75}. A fascinating property from a purely mathematical point of view is that there 
is no dominance of either of the two terms $r^2$ and $\lambda/r^{\alpha}$ for the extreme values 
of $\lambda$. In other words, there is a tug-of-war between the two terms. Typical shape of the 
potential for a few parameter sets are depicted in Fig. 1

Like most of the practical physical systems, exact analytical solution of the corresponding 
Schr\"odinger equation (SE) for arbitrary parameters for a general state (both ground and excited) 
is yet to be known and consequently approximations such as variational or perturbation methods 
must be used. Thus a large number of attempts have been made towards the exact and approximate 
calculation of eigenvalues and eigenfunctions employing analytic, seminumerical or purely 
computational methods [3-18]. As the Rayleigh-Schr\"odinger perturbation series for the eigenvalues 
of the operator $H$ diverges \cite{harrell77}, a modified perturbation theory to finite order was
developed which gave good result for the ground-state eigenvalues of small 
$\lambda$. For large positive values of $\lambda$, approximate estimates were made through the 
large coupling perturbative expansion \cite{harrell77}. A resummation technique for weak coupling 
of the nonsingular SHO ($\alpha<5/2$) \cite{navarro90} was developed. Exact and approximate 
solutions are also given for the lowest states of a given angular momentum \cite{fernandez91}. A 
WKB treatment has been proposed \cite{trost97}. Recently extensive studies have been made on 
the upper and lower bounds as well as the direct eigenvalues by means of the geometrical 
approximation and method of potential envelopes \cite{hall95,hall98,hall02}. The analytical 
pseudoperturbation shifted-$\ell$ expansion \cite{mustafa99,mustafa00} has also been suggested.
Reliable results were reported by using a Richardson extrapolation of the finite-difference scheme 
\cite{killingbeck01}, integrating the radial Schr\"odinger equation by Lanczos grid method 
\cite{torres92}, analytic continuation method \cite{buendia95}, etc. Recently a generalized 
pseudospectral (GPS) method was shown to produce very accurate results for the SHO eigenvalues, 
eigenfunctions and other properties \cite{roy04}. Arbitrary eigenstates corresponding to any given
$n$ and $\ell$ values were treated for any general potential parameters within a simple and unified 
manner. 
 
In parallel to these above works, recently there has been significant interest [19-24] in studying a 
\emph{generalized spiked harmonic oscillator} (GSHO) as well, defined by the Hamiltonian, 
\begin{equation}
H= -\frac{d^2}{dr^2}+v(r); \ \ \ v(r)=r^2+\frac{A}{r^2}+\frac{\lambda}{r^{\alpha}}, \ \ \ A \ge 0
\end{equation}
both using variational and perturbation methods. Here $\lambda$, $\alpha$ are two real parameters 
and the SHO is a special case of GSHO with $A=0$. The present study is motivated by these recent
works of Hall and coworkers [19-24]. In \cite{saad03} using a perturbation expansion up to 3rd order, 
they estimated the \emph{lower and upper bounds} of the GSHO ground states for \emph{small} coupling 
parameter $\lambda$ (four values in the range of 0.001--1). However no \emph{direct} results were 
given. Other than this, we are not aware of any
other works towards the calculation of these bound states or even the approximate bounds. Hence it 
would be of some interest to investigate the spectra of this system in some detail. The objective 
of this work is thus to report the accurate eigenvalues, eigenfunctions, expectation values of some 
position operators of the GSHO for the first time. As shown in later sections, both \emph{lower} and 
\emph{higher} states can be tackled
very efficiently for arbitrary parameters including both \emph{weak} and \emph{strong} coupling in a 
simple manner. Moreover, in order to understand the spectra, we make a detailed analysis on the effect of 
variation of the parameters on eigenvalues and radial densities. For this we employ the GPS method,
which has been very successful to produce accurate and reliable results in recent years for a variety of 
singular potentials relevant to quantum mechanics, static and dynamic properties of atom, molecules, 
as well as the spherically confined isotropic harmonic oscillator [18,25-28]. The article is organized 
as follows: Section II gives a brief outline of the GPS method used here to solve the SE for the GSHO. 
A discussion of the results is made in Section III, while we make a few concluding remarks in 
Section IV.    

\begin{figure}
\centering
\includegraphics[scale=0.4]{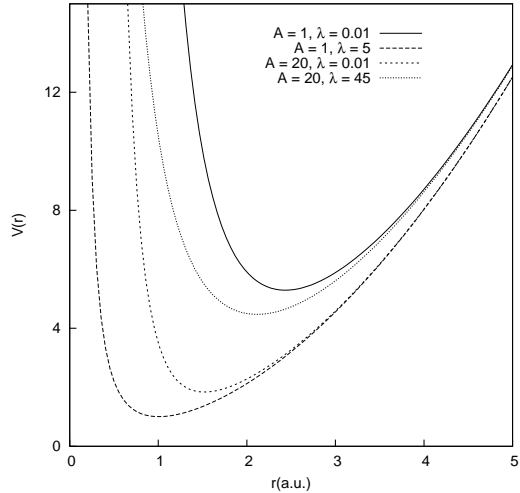}
\caption{The potential $V(r)=v(r)/2$, for $\alpha=4$ for several values of $A$ and
$\lambda$.} 
\end{figure}

\section{Method of Calculation}
\label{sec:method}
This section presents an overview of the GPS formalism used for solving the radial SE of a 
Hamiltonian containing a GSHO potential within the nonrelativistic framework. Only essential steps 
are given as relevant details have been presented earlier [18,25-28]. Unless otherwise mentioned, 
atomic units are employed throughout this Letter. 

The radial SE can be written in the following form, 
\begin{equation}
\left[-\frac{1}{2} \ \frac{\mathrm{d^2}}{\mathrm{d}r^2} + \frac{\ell (\ell+1)} {2r^2}
+V(r) \right] \psi_{n,\ell}(r) = E_{n,\ell}\ \psi_{n,\ell}(r)
\end{equation}
where $V(r)=v(r)/2$ and $v(r)$ is as given in Eq. (1). Note that a 2 factor is introduced for 
the sake of consistency with the literature. Here $n$ and $\ell$ signify the usual radial and angular 
momentum quantum numbers respectively. 

One of the distinctive features of GPS method is that it allows one to work in a nonuniform and 
optimal spatial discretization; a coarser mesh at larger $r$ and a denser mesh at smaller $r$, 
while maintaining a similar accuracy at \emph{both} the regions. Thus it suffices to work with a 
significantly smaller number of radial points efficiently, which is in sharp contrast to some of the 
commonly used finite difference or finite element methods for the singular potentials, where one is 
almost forced to use considerably larger mesh, often presumably because of their uniform nature. 
This feature was realized and exploited during the recent accurate calculation of high-lying 
multiply excited \emph{Rydberg} series of Li isoelectronic series \cite{roy04a} using 
density functional theory. 

The key step is to approximate a function $f(x)$ defined in the interval $x \in [-1,1]$ by the N-th 
order polynomial $f_N(x)$ as follows through a cardinal function $g_j(x)$,
\begin{equation}
f(x) \cong f_N(x) = \sum_{j=0}^{N} f(x_j)\ g_j(x),
\end{equation}
which guarantees that the approximation is \emph {exact} at the \emph {collocation points} $x_j$, 
i.e., $ f_N(x_j) = f(x_j)$, requiring that the cardinal function satisfies $g_j(x_{j'})=\delta_{j'j}$.
In the Legendre pseudospectral method which we use here, $x_0=-1$, $x_N=1$, and the $x_j 
(j=1,\ldots,N-1)$ are obtained from the roots of the first derivative of the Legendre polynomial 
$P_N(x)$ with respect to $x$ as $ P'_N(x_j) = 0$.
The $g_j(x)$ are given by,
\begin{equation}
g_j(x) = -\frac{1}{N(N+1)P_N(x_j)}\ \  \frac{(1-x^2)\ P'_N(x)}{x-x_j},
\end{equation}
Now one can map the semi-infinite domain $r \in [0, \infty]$ onto the finite domain $x \in [-1,1]$ 
by the transformation $r=r(x)$. At this stage, introduction of an algebraic nonlinear
mapping, 
\begin{equation} 
r=r(x)=L\ \frac{1+x}{1-x+\gamma},
\end{equation}
where L and $\gamma=2L/r_{max}$ are the mapping parameters, 
gives rise to the transformed differential equation as: $ f(x)=R_{nl}(r(x))/\sqrt{r'(x)}$. Applying 
the Legendre pseudospectral method now to this equation followed by a symmetrization procedure 
finally leads to a \emph{symmetric} eigenvalue problem,
\begin{equation}
\sum_{j=1}^{N-1} \left[ -\frac{1}{2}D_{ij}+u_j\delta_{ij} \right] \chi_j = \epsilon_{nl} \chi_i
\end{equation}
which can be easily solved by standard available routines to yield highly accurate eigenvalues 
and eigenfunctions. Here 
\begin{equation}
\chi_i\!= \! R_{nl}(r_i) \ \sqrt{(r'_i)}/P_N(x_i), \ \ \ u_i\!=\!l(l+1)/2r_i^2 
+v(r_i)
\end{equation}
with $\chi_i\!=\!\chi(x_i), u_i\!=\!u(x_i), r_i\!=\!r(x_i), r'_i\!=r'(x_i)$, and $D_{ij}$ 
denoting the symmetrized second derivative of the cardinal function given as [18,25-28]. 
\begin{eqnarray}
D_{ij} & = & -\frac{2}{r_i'(x_i-x_j)^2r'_j}, \  \ \ \ i \neq j, \nonumber \\
& = & -\frac{N(N+1)}{3r_i'^2(1-x_i^2)}, \ \ \ i= j.
\end{eqnarray}
A large number of tests were carried out in order to make a detailed check on the accuracy and 
reliability of our calculation by varying the mapping parameters so as to produce ``stable'' results.
In this way, a consistent set 
of parameters $\gamma=25, N=200$ and $r_{max}=300$ were chosen which seemed to be appropriate for 
all the calculations performed in this work. The results are reported only up to the precision that 
maintained stability. The energy values are {\em truncated} rather than {\em rounded-off} and thus 
may be considered as correct up to the place they are reported. 

\section{Results and Discussion}
First in Table I we give a comparison of the GPS eigenvalues of GSHO with the available literature 
data as an illustration to assess the validity and usefulness of our present calculation. Lowest two 
eigenvalues of $\ell=0,1,2,3$ are given at six values of $\lambda$ in the range of 0.001--100 covering 
both weaker and stronger interaction to demonstrate the generality of our results. The only result
available in the literature corresponds to $A=12$, $\alpha=4$ and we stick to them, although we are 
confident that this will produce equally correct results for any other set of parameters 
or states. The lowest state corresponding to $\ell=0$ for $\lambda=0.001,0.1,0.1$ and 1 have been 
studied through the upper and lower bounds \cite{saad03}, and those are presented in the footnote. Note 
that those results are divided by 2 to get the entries in the table. They used a third-order perturbation 
expansion using the procedure suggested by Burrows {\emph et al} \cite{burrows87}. As pointed out in 
\cite{saad03}, there are a number of variational methods available for the GSHO Hamiltonian, which can 
provide accurate \emph{upper} bounds in question, but they do not lead to \emph{direct} evaluation of 
energies. Furthermore, for \emph{smaller} $\lambda$ values, these methods are slow and in general, a 
large number of matrix elements are required to achieve a reasonable accuracy. As can be seen from the 
table, our results give estimates for all the four values of $\lambda$ that perfectly lie in between 
those bounds. Clearly as anticipated, for smaller $\lambda$ values the bounds are narrow and hence
very accurate 
methods such as the present ones would be needed to generate correct values. No results are available for 
any other parameter sets or states and our calculation could be helpful in future studies of these 
systems. We also did similar calculations for $\alpha=6$ with $A=12$ leading to similar kind of results 
and conclusions and thus are omitted. 

\begingroup
\squeezetable
\begin{table}
\caption {\label{tab:table1}Comparison of the calculated GSHO eigenvalues (in a.u.) for $A=12$ and
$\alpha=4$. First two eigenvalues are given for each $\ell$. Literature results are appropriately 
converted to the current scale of units.}
\begin{ruledtabular}
\begin{tabular}{ccccc}
$\lambda$  &  $\ell = 0$       & $\ell=1$            & $\ell=2$         & $\ell=3$        \\   \hline
0.001      & 4.50005713955\footnotemark[1]  & 4.77496494795   & 5.27203764224  & 5.92445477296   \\
           & 6.50008253170     & 6.77498493821       & 7.27205121112    & 7.92446350670   \\
0.01       & 4.50057109969\footnotemark[2]  & 4.77539434151       & 5.27235948689    & 5.92468758726   \\
           & 6.50082460262     & 6.77559400403       & 7.27249507610    & 7.92477488725   \\
0.1        & 4.50568201308\footnotemark[3]  & 4.77967076076       & 5.27556990359    & 5.92701233962   \\
           & 6.50817676852     & 6.78164398019       & 7.27691597471    & 7.92788162240   \\
1          & 4.55432930375\footnotemark[4]  & 4.82086660209       & 5.30692177482    & 5.94993288816   \\
           & 6.57618592946     & 6.83867710911       & 7.31951196183    & 7.95827867190   \\
10         & 4.91961566042     & 5.14553963970       & 5.57091626978    &  6.15427959467  \\
           & 7.03453114315     & 7.24781484749       & 7.65332837563    &  8.21617552368  \\
100        & 6.54544524211     & 6.68956373653       &  6.9715527505    &  7.37986452167  \\
           & 8.81244551008     & 8.94633479206       &  9.2093889489    &  9.59269153472  \\
\end{tabular}
\end{ruledtabular}
\footnotetext[1] {The lower and upper bounds are 4.5000571155 and  4.5000571635 \cite{saad03}.}
\footnotetext[2] {The lower and upper bounds are 4.5005687045 and  4.5005734945 \cite{saad03}.}
\footnotetext[3] {The lower and upper bounds are 4.5054425485 and  4.5059217125 \cite{saad03}.}
\footnotetext[4] {The lower and upper bounds are 4.5306226410 and  4.5786886205 \cite{saad03}.}
\end{table}
\endgroup

Next we consider the variation of eigenvalues of the GSHO through a cross-section of the sampled 
results for a varied range of representative parameter sets and states in Table II. For this, we choose
four values of 
$\lambda$ lying within the broad range of 0.005--50. For each $\lambda$, three values of $A$ are selected, 
namely 5, 15 and 25 respectively. The angular quantum momentum $\ell$ is chosen at random from 0--3 and 
first two states belonging to each $\ell$ are given. We restrict ourselves to only two $\alpha$ values of
4 and 6, corresponding to the left and right panel respectively. No results have been reported so far in 
the literature for direct comparison and we hope these will inspire further works in future.  
 
\begingroup
\squeezetable
\begin{table}
\caption {\label{tab:table2}The calculated eigenvalues (in a.u.) of GSHO for various parameters.
Left for $\alpha =4$, right for $\alpha= 6$. Lowest two energies are given for each $\ell$.}
\begin{ruledtabular}
\begin{tabular}{ccccccc}
$\lambda$  & A   & $\ell $  & \multicolumn{2}{c}{$\alpha =4$}  & \multicolumn{2}{c}{$\alpha =6$} \\
\cline{4-5}     \cline{6-7} 
        &    &     &      $E_1$      &    $E_2$        &   $E_1$          &  $E_2$          \\  \hline
 0.005  & 5  & 0   &  3.29213042081  &  5.29263961857  &  3.29301057259   & 5.29560957327   \\
        & 15 & 0   &  4.90534516173  &  6.90543495986  &  4.90524031751   & 6.90538113434   \\
        & 25 & 0   &  6.02506141110  &  8.02510243449  &  6.02497866843   & 8.02501934461   \\
 0.5    & 5  & 1   &  3.74338802444  &  5.76740985617  &  3.73736635296   & 5.78462712272   \\
        & 15 & 1   &  5.17214026739  &  7.17923504322  &  5.16180870804   & 7.17100250031   \\
        & 25 & 1   &  6.23143489376  &  8.23501549308  &  6.22364595141   & 8.22695053203   \\
    5   & 5  & 2   &  4.60929429358  &  6.69135284445  &  4.49045054121   & 6.59803098276   \\
        & 15 & 2   &  5.74836543195  &  7.79050612724  &  5.66024910570   & 7.70391813910   \\
        & 25 & 2   &  6.68350815067  &  8.70939627752  &  6.61603417371   & 8.63746955793   \\
   50   & 5  & 3   &  6.25457661591  &  8.44871213892  &  5.57496724850   & 7.79410043287   \\
        & 15 & 3   &  7.03078249009  &  9.18083908970  &  6.46642800655   & 8.61050093425   \\
        & 25 & 3   &  7.74063452037  &  9.85964537843  &  7.26355121587   & 9.36136024083   \\
\end{tabular}
\end{ruledtabular}
\end{table}
\endgroup

The above variation of energies with potential parameters are shown more clearly in Fig.~2. The top panel 
(a)--(c) depicts the $\lambda$ vs. $E$ plot for three $A$ values, \emph{viz.,} 5, 15, 25 respectively. For 
each $A$, the first three states of $\ell=0$ are presented ($E_1$, $E_2$, $E_3$ from bottom to top) and 
$\lambda$ varied from 0--35. These are studied for both $\alpha=4$ and 6 as marked in the figure. Note
that the energy axes in (a)--(c) and (d)--(f) are kept fixed at 0--12 and 2--13. The 
variation of energies with respect to $\lambda$ are rather slow (monotonically increases as $\lambda$ 
increases) for both $\alpha=4$ and 6, with smaller $\alpha$ exerting relatively larger effect. At 
smaller $\lambda$ (in the vicinity of zero), the $\alpha=4$ and 6 energies are quite close to each other
for all values of $A$
and with an increase in $\lambda$ they show pronounced separation. As $A$ increases, the slope of 
$\lambda$ vs. $E$ decreases, making the plot rather flat as in (c), signifying even slower variation of 
$E$ with respect to $\lambda$. In the bottom, we show the $A$ vs. $E$ in the range of $A=5-50$ for three 
values of $\lambda$, namely 1, 10, 25 respectively in (d)--(f). Once again we show the first three states 
of $\ell=0$ of both $\alpha =4$ and 6, which make three separate families. For a given $\lambda$, as $A$ 
increases, $E$ increases rather sharply compared to those in top panel, (a)--(c). Interestingly however, 
for $\lambda=1$, the $\alpha=4$ and 6 plots seem to be virtually identical for all the three states. As 
$\lambda$ increases the separation gradually increases slowly as one passes from (d)-(e)-(f).  

\begin{figure}
\centering
\begin{minipage}[t]{0.27\textwidth}\centering
\includegraphics[scale=0.32]{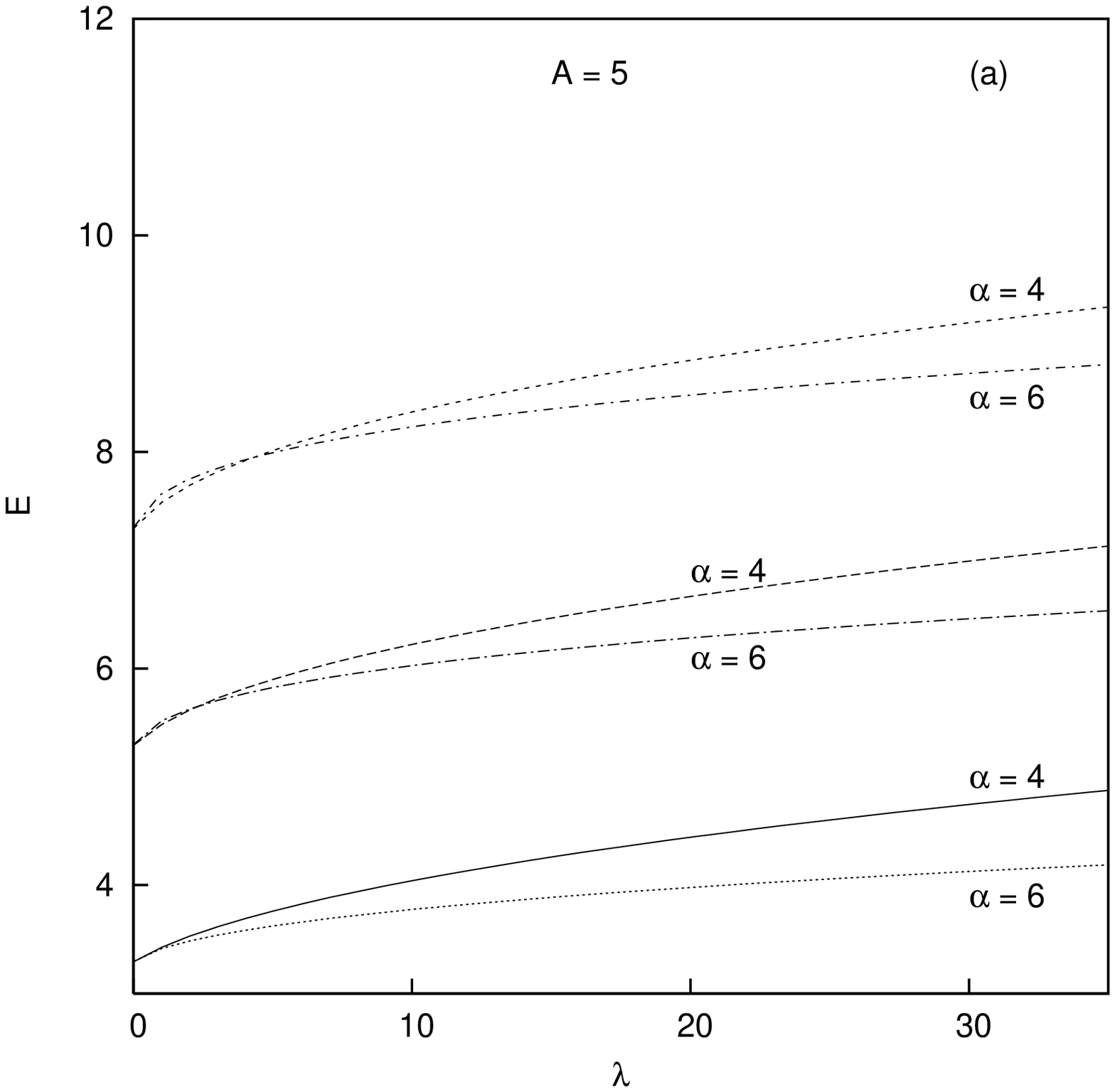}
\end{minipage}
\hspace{0.40in}
\begin{minipage}[t]{0.27\textwidth}\centering
\includegraphics[scale=0.32]{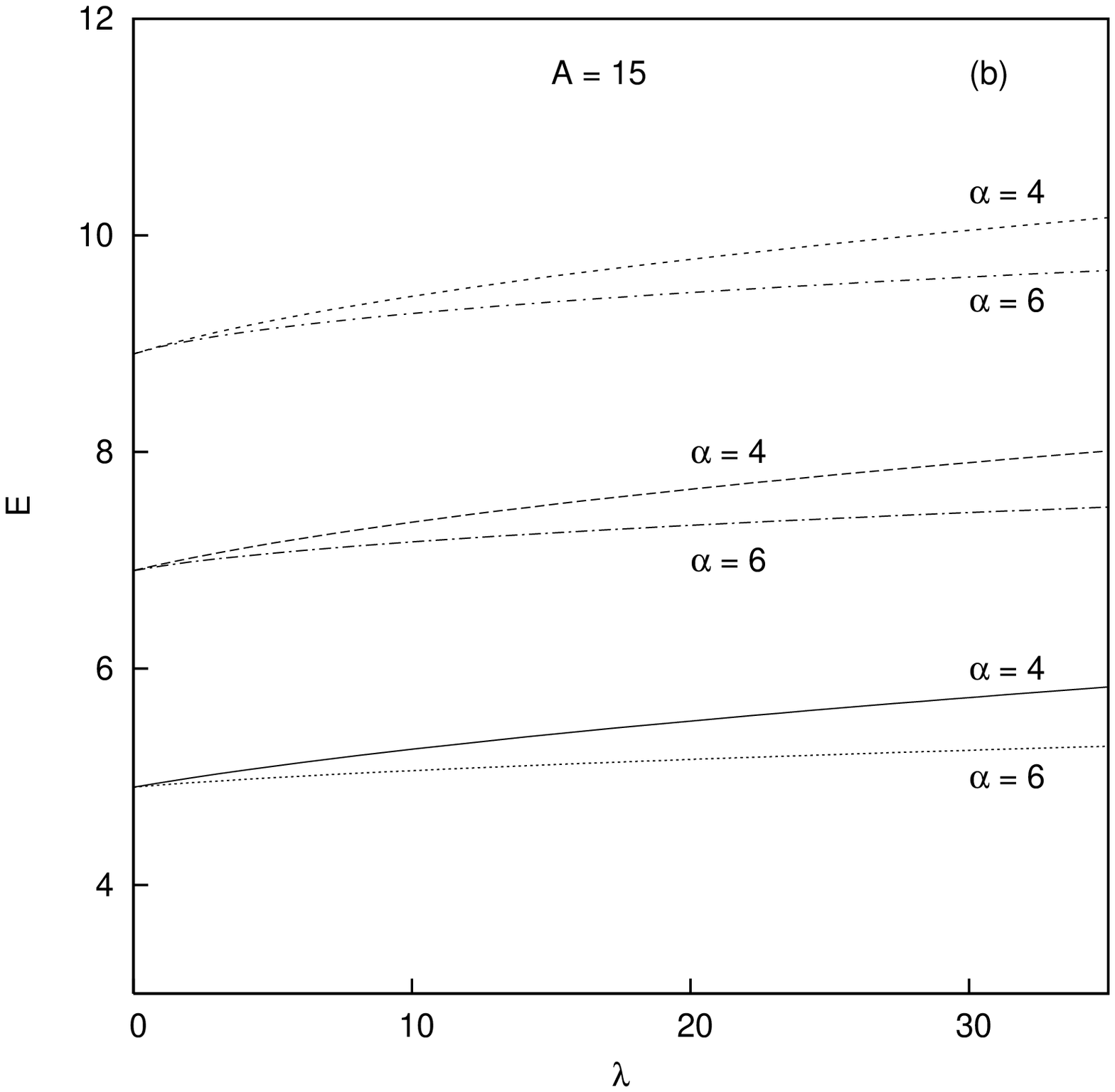}
\end{minipage}
\hspace{0.40in}
\begin{minipage}[t]{0.27\textwidth}\centering
\includegraphics[scale=0.32]{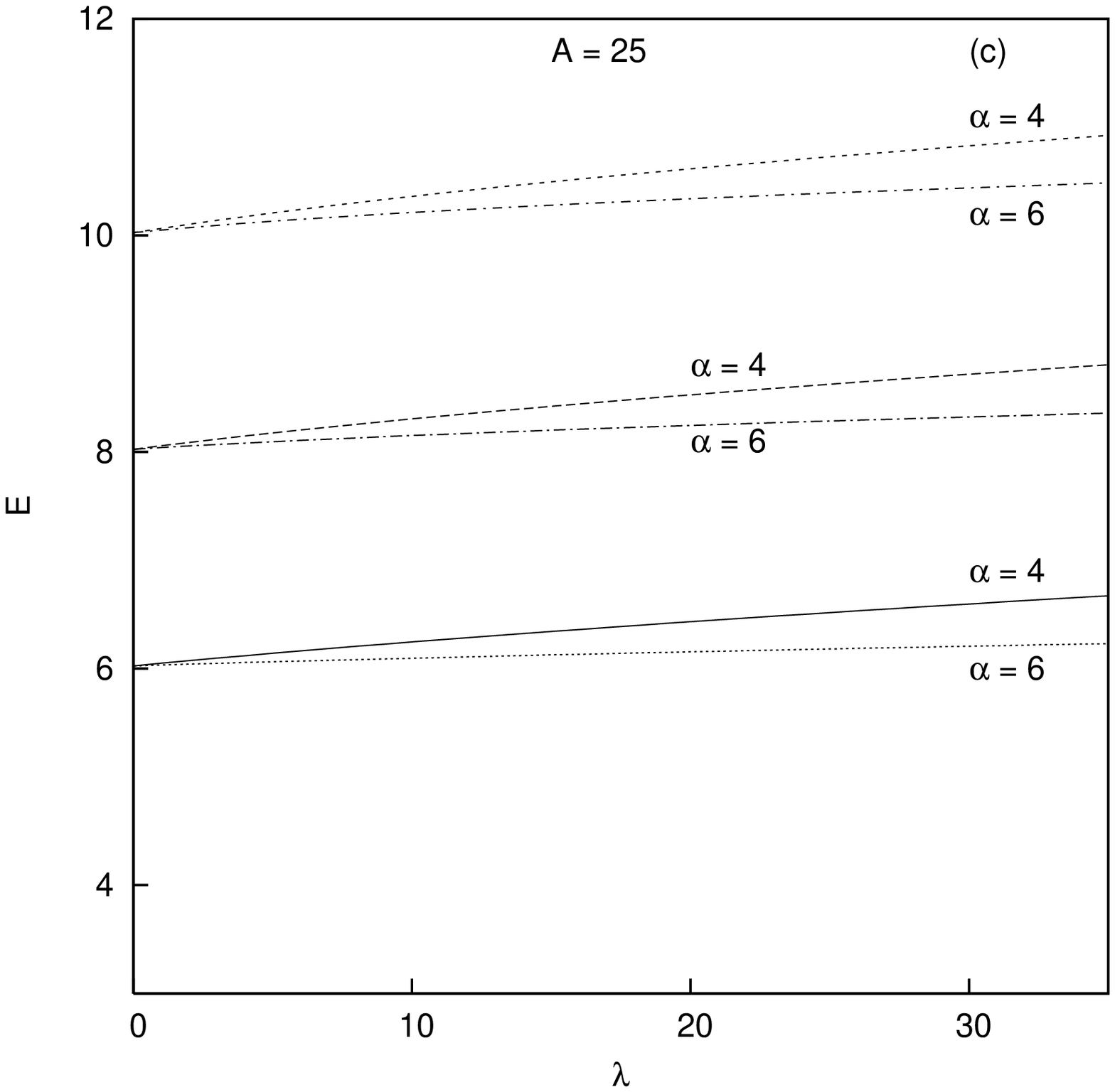}
\end{minipage}
\\[30pt]
\begin{minipage}[b]{0.27\textwidth}\centering
\includegraphics[scale=0.32]{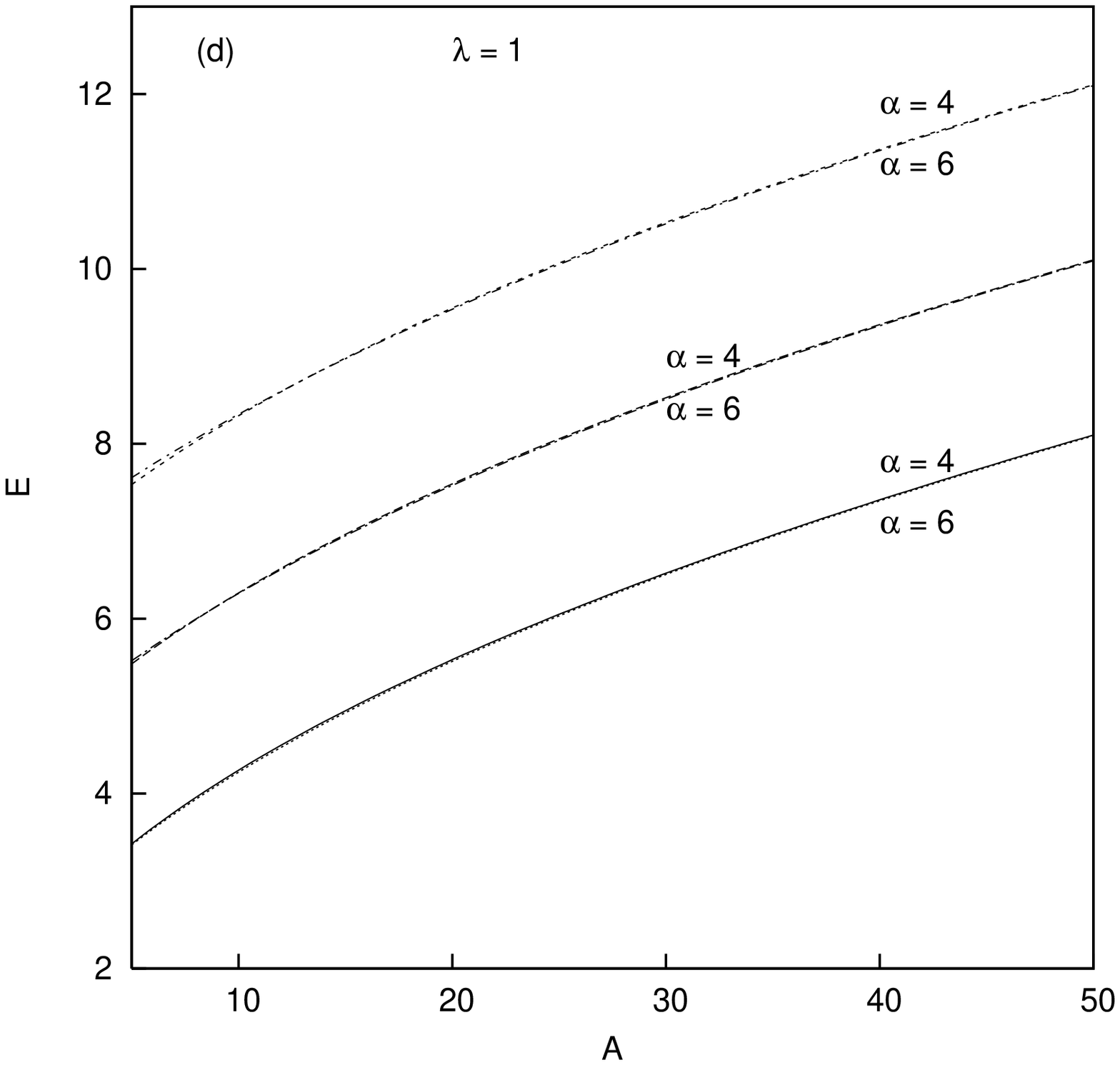}
\end{minipage}
\hspace{0.40in}
\begin{minipage}[b]{0.27\textwidth}\centering
\includegraphics[scale=0.32]{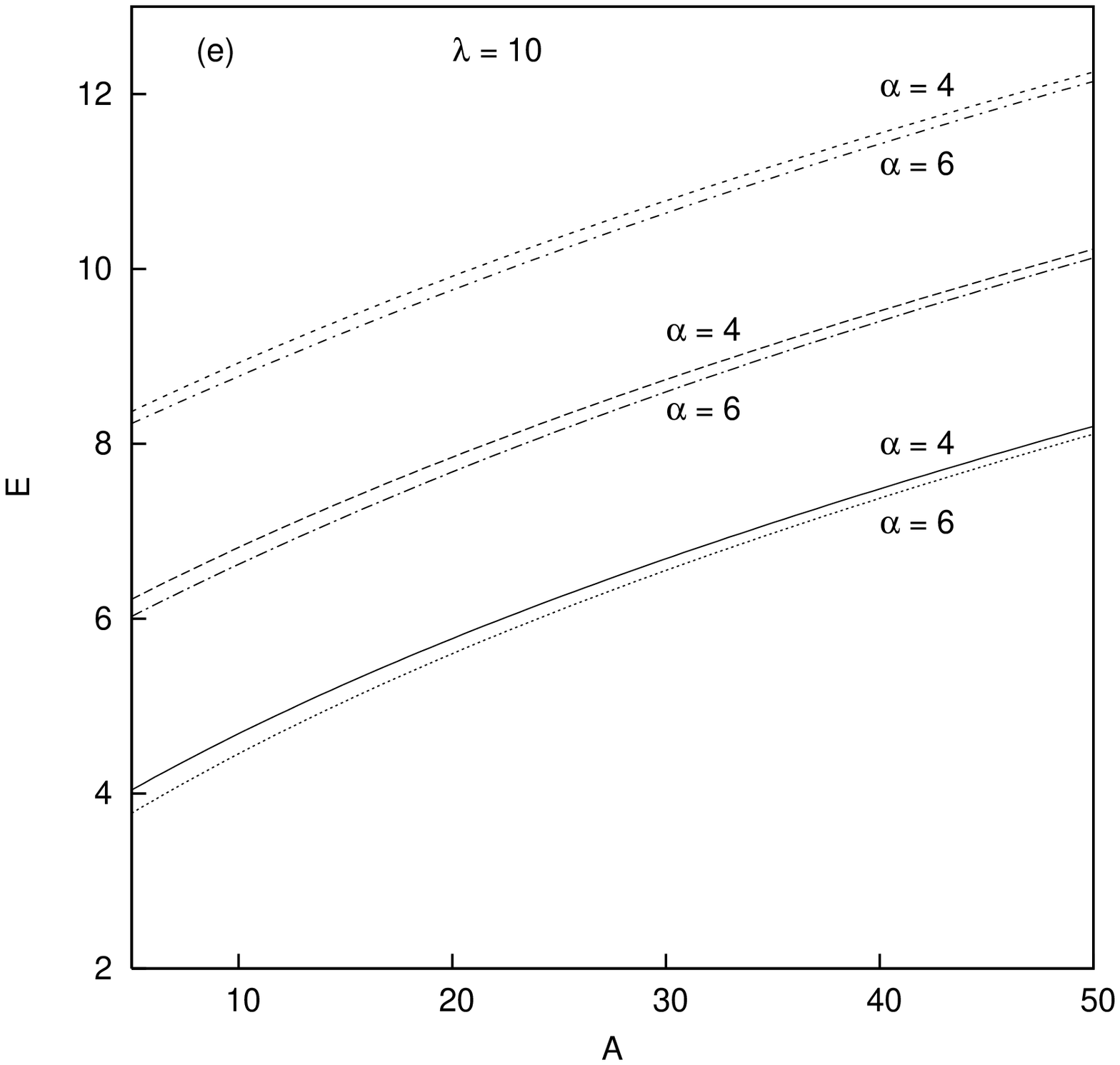}
\end{minipage}
\hspace{0.40in}
\begin{minipage}[b]{0.27\textwidth}\centering
\includegraphics[scale=0.32]{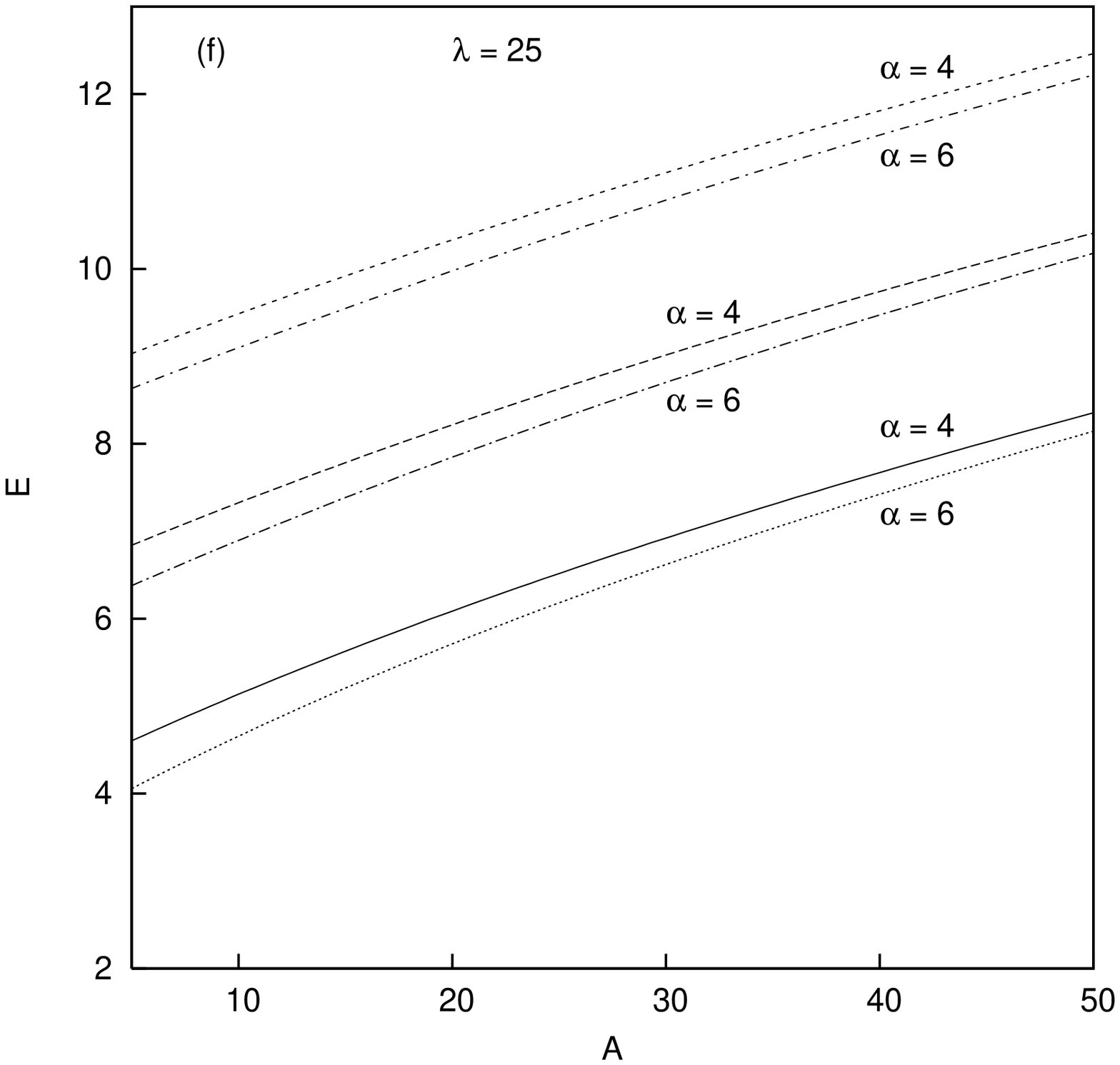}
\end{minipage}
\caption[optional]{Variation of $E$ with respect to $\lambda$ (top) and $A$ (bottom) respectively for 
the first three states belonging to $\ell=0$ of a GSHO having $\alpha=4$ and 6. (a) $A=5$, (b) $A=15$, 
(c) $A=25$; (d) $\lambda=1$, (e) $\lambda=10$, (f) $\lambda=25$.}
\end{figure}

Next in Fig.~3 we show the changes in ground-state radial probability distribution function 
$|rR_{n,l}|^2$ with respect to the potential parameters $\lambda$ (a) and $A$ (b) respectively. In (a) 
we fix $A$ at 20 and four plots are given for $\lambda=0.01, 50, 200$ and 500. With an increase in 
$\lambda$, the peak position and peak height shifts to higher values. In (b) we show the variation at 
four values of $A$ (1, 10, 20 and 30) by keeping $\lambda$ constant at 1. In this case, with an increase
in $A$, the peak position increases, but the peak height decreases.  

\begin{figure}
\centering
\begin{minipage}[t]{0.4\textwidth}\centering
\includegraphics[scale=0.39]{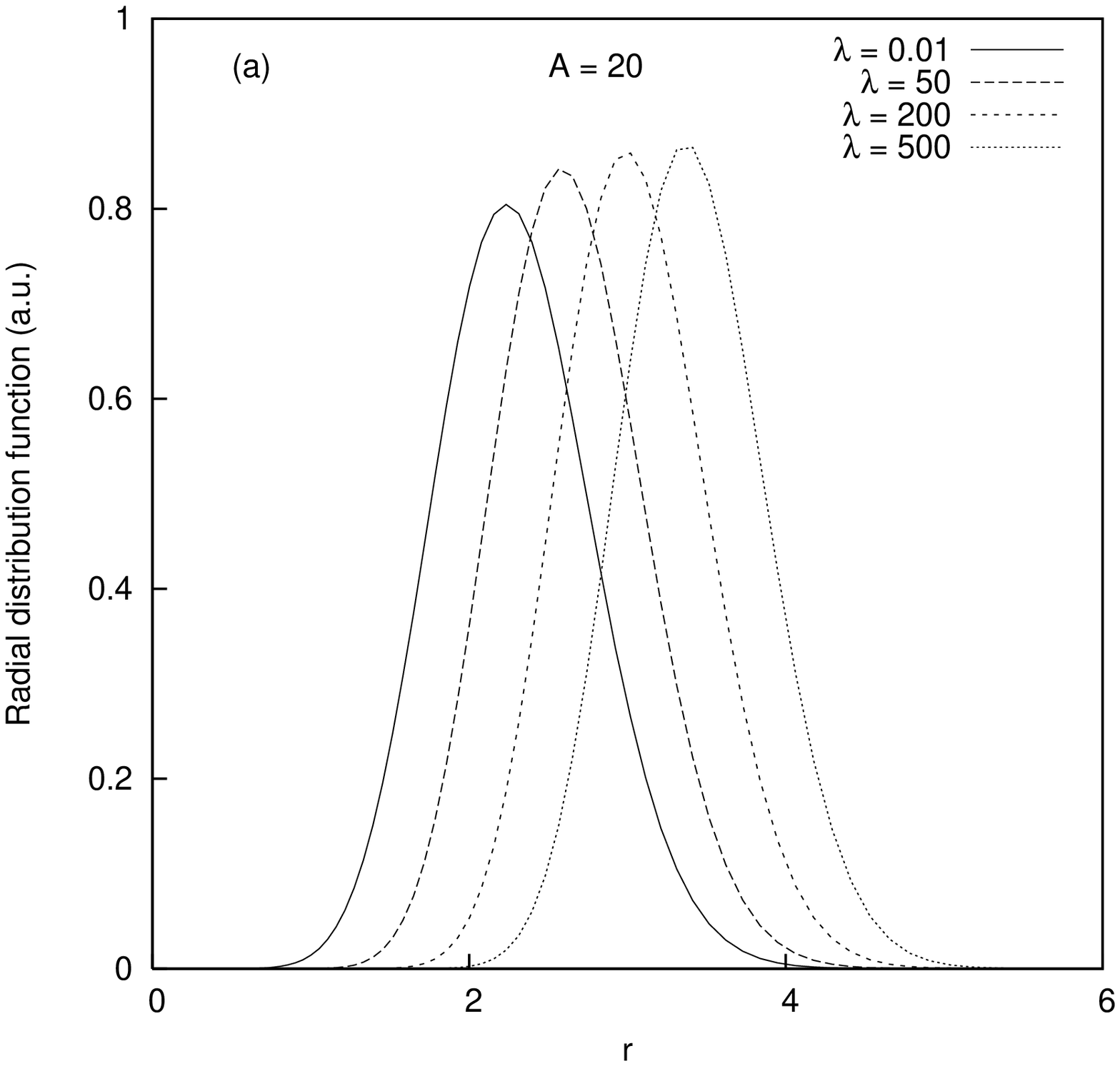}
\end{minipage}
\hspace{0.15in}
\begin{minipage}[t]{0.35\textwidth}\centering
\includegraphics[scale=0.39]{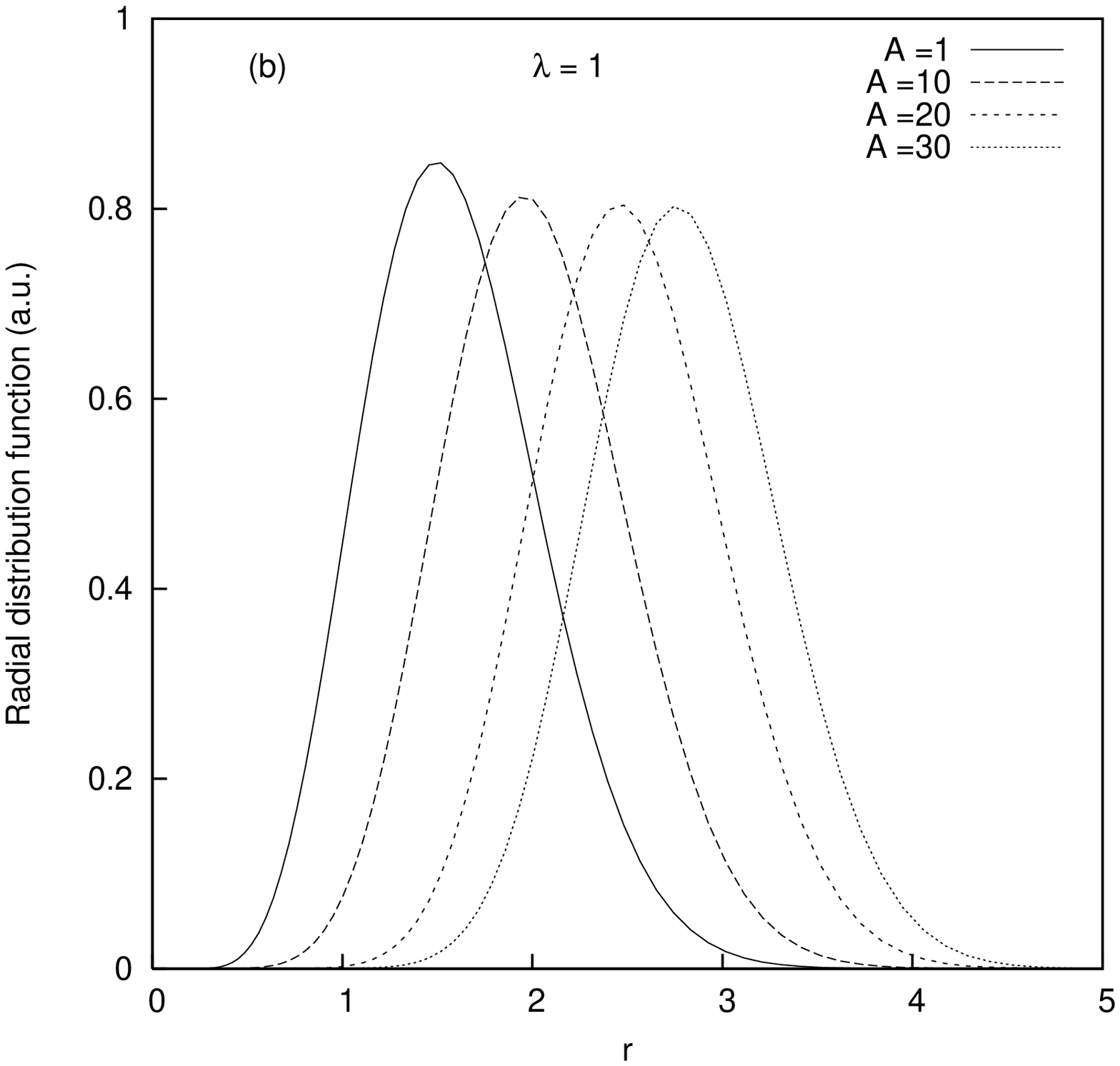}
\end{minipage}
\caption[optional]{Change in  radial probability distribution function, $|rR_{n,\ell}|^2$, for the
ground state of GSHO with $\lambda$ and A. (a) $\lambda=0.001,50,200 $ and 500 at $A=20$,
(b) $A=1,10,20,30$ at $\lambda=1$.}
\end{figure}

As a demonstration of the accuracy of our calculated wave functions, finally in Table III, two 
position expectation values $\langle r^{-1}\rangle$ and  $\langle r \rangle $ (in a.u.) of the GSHO with 
$\lambda=1$ and 10 are reported for $\alpha=4$ (left) and 6 (right) respectively. First two states of 
$\ell=0$ are given keeping $A$ fixed at 20. There has been no result of these values for comparison.

\begingroup
\squeezetable
\begin{table} 
\caption {\label{tab:table3}Calculated expectation values (in a.u.) for the GSHO for some selected values 
of $\lambda$ for $\alpha=4$ (left) and 6 (right). First two states belonging to $\ell=0 $ are presented. 
$A$ = 20.} 
\begin{ruledtabular}
\begin{tabular} {cccccccc}
\multicolumn{4}{c}{$\alpha =4$} &  \multicolumn{4}{c}{$\alpha=6$}   \\
\cline{1-4}  \cline{5-8} 
$\lambda$ &$\ell$ &$\langle r^{-1}\rangle$  & $\langle r \rangle $  &
$\lambda$ &$\ell$ &$\langle r^{-1}\rangle$  & $\langle r \rangle $ \\ \hline 
  1       &   0   & 0.455313939   & 2.30630576  &   1   &  0   & 0.456468262 &  2.300730227  \\     
          &       & 0.433586990   & 2.62193026  &       &      & 0.433801832 &  2.619954805  \\
   10     &   0   & 0.434300669   & 2.40340664  &   10  &  0   & 0.444230432 &  2.353215494  \\
          &       & 0.409762190   & 2.72989436  &       &      & 0.415474164 &  2.697640059  \\
\end{tabular}
\end{ruledtabular}
\end{table}
\endgroup

\section{Conclusion}
Discrete bound-state spectra of the GSHO are studied for the first time in detail by accurately 
calculating the eigenvalues, eigenfunctions, position expectation values and densities by means of the 
GPS method. The formalism is quite simple, computationally efficient, reliable and as illustrated,
very accurate. Low as well as high states are calculated for arbitrary values of the interaction 
parameters covering weak and strong couplings with equal ease and accuracy. Only estimates of the upper 
and lower bounds of a few ground states have been reported so far in the literature; no direct results
are available for the GSHO, not to speak of the highly accurate values. Thus we hope that our 
calculation of these  hitherto unreported results would constitute a useful reference for future
works on this system. In addition, the variation of eigenvalues and densities with respect to the 
potential parameters are also discussed. 

\begin{acknowledgments}
AKR thanks Professors D.~Neuhauser and S.~I.~Chu for encouragement, financial support and useful 
discussions. He thanks Dr.~E.~I.~Proynov for helpful discussion. AKR acknowledges 
the warm hospitality provided by Univ. of California, Los angeles, CA, USA.
\end{acknowledgments}

\end{document}